%% file: sample-sigconf.tex
\documentclass[sigconf]{acmart}

\usepackage{courier}  
\usepackage{graphicx} 
\usepackage{natbib}  
\usepackage{xcolor}
\usepackage{caption} 
\newcommand{\xhdr}[1]{{\vspace{1pt}\noindent\bfseries #1}.}
\newcommand{\ie}{\textit{i.e., }}
\newcommand{\eg}{\textit{e.g., }}

\usepackage{multirow}
\usepackage{xspace}
\usepackage{subfigure}
\newcommand{\method}{LitFM\xspace}
\newcommand{\argtopk}{\mathop{\mathrm{arg\,top}\,k}}
\usepackage{algorithm}
\usepackage{tabularx}
\usepackage{booktabs}
\usepackage{arydshln} 
\usepackage{nicematrix}
\usepackage{algorithmic}
\usepackage{amsfonts, amsmath}
\usepackage{enumitem}
\definecolor{niceblue}{HTML}{007ED6}

\usepackage{newfloat}
\usepackage{listings}
\AtBeginDocument{%
  }

\setcopyright{acmlicensed}
\copyrightyear{2018}
\acmYear{2018}
\acmDOI{XXXXXXX.XXXXXXX}
\acmConference[Conference acronym 'XX]{Make sure to enter the correct
  conference title from your rights confirmation email}{June 03--05,
  2018}{Woodstock, NY}
\acmISBN{978-1-4503-XXXX-X/2018/06}

\begin{document}

\title{\method: A Retrieval Augmented Structure-aware Foundation Model For Citation Graphs}

\author{Jiasheng Zhang}
\affiliation{
\institution{University of Electronic Science and Technology of China}
\country{Chengdu, China}
}
\email{zjss12358@std.uestc.edu.cn}

\author{Ali Maatouk}
\affiliation{
\institution{Yale University}
\country{New Haven, USA}
}
\email{ali.maatouk@yale.edu}

\author{Jialin Chen}
\affiliation{
\institution{Yale University}
\country{New Haven, USA}
}
\email{jialin.chen@yale.edu}

\author{Ngoc Bui}
\affiliation{
\institution{Yale University}
\country{New Haven, USA}
}
\email{ngoc.bui@yale.edu}

\author{Qianqian Xie}
\affiliation{
\institution{Yale University}
\country{New Haven, USA}
}
\email{qianqian.xie@yale.edu}

\author{Leandros Tassiulas}
\affiliation{
\institution{Yale University}
\country{New Haven, USA}
}
\email{leandros.tassiulas@yale.edu}

\author{Hua Xu}
\affiliation{
\institution{Yale University}
\country{New Haven, USA}
}
\email{hua.xu@yale.edu}

\author{Jie Shao*}
\affiliation{
\institution{University of Electronic Science and Technology of China}
\country{Chengdu, China}
}
\email{shaojie@uestc.edu.cn}

\author{Rex Ying}
\affiliation{
\institution{Yale University}
\country{New Haven, USA}
}
\email{rex.ying@yale.edu}

\thanks{$^*$Corresponding author: Jie Shao.}
\renewcommand{\shortauthors}{Jiasheng Zhang et al.}

\input{chapter/00_abstract}




\maketitle

\input{chapter/01_intro}

\input{chapter/02_relatedwork}

\input{chapter/03_method}
\input{chapter/04_experiment}

\bibliographystyle{ACM-Reference-Format}
\bibliography{sample-base}

\appendix
\input{chapter/05_appendices}

\end{document}

%% file: chapter/00_abstract.tex
\begin{abstract}

With the advent of large language models (LLMs), managing scientific literature via LLMs has become a promising direction of research. However, existing approaches often overlook the rich structural and semantic relevance among scientific literature, limiting their ability to discern the relationships between pieces of scientific knowledge, and suffer from various types of hallucinations. These methods also focus narrowly on individual downstream tasks, limiting their applicability across use cases. Here we propose \method, the first literature foundation model designed for a wide variety of practical downstream tasks on domain-specific literature, with a focus on citation information. At its core, \method contains a novel graph retriever that can provide accurate and diverse recommendations for LLM to integrate graph structure information and relevant literature. \method also leverages a knowledge-infused LLM, fine-tuned through a well-developed instruction paradigm. It enables \method to extract domain-specific knowledge from literature and reason relationships among them. By integrating citation graphs during both training and inference, \method can generalize to unseen papers and accurately assess their relevance within existing literature. Additionally, we introduce new large-scale literature citation benchmark datasets on three academic fields, featuring sentence-level citation information and local context. Extensive experiments validate the superiority of \method, achieving $28.1\%$ improvement on retrieval task in precision, and an average improvement of $7.52\%$ over state-of-the-art across six downstream literature-related tasks. 

\end{abstract}

%% file: chapter/01_intro.tex

\section{Introduction}
\begin{figure}
    \centering
    \includegraphics[width=0.85\linewidth]{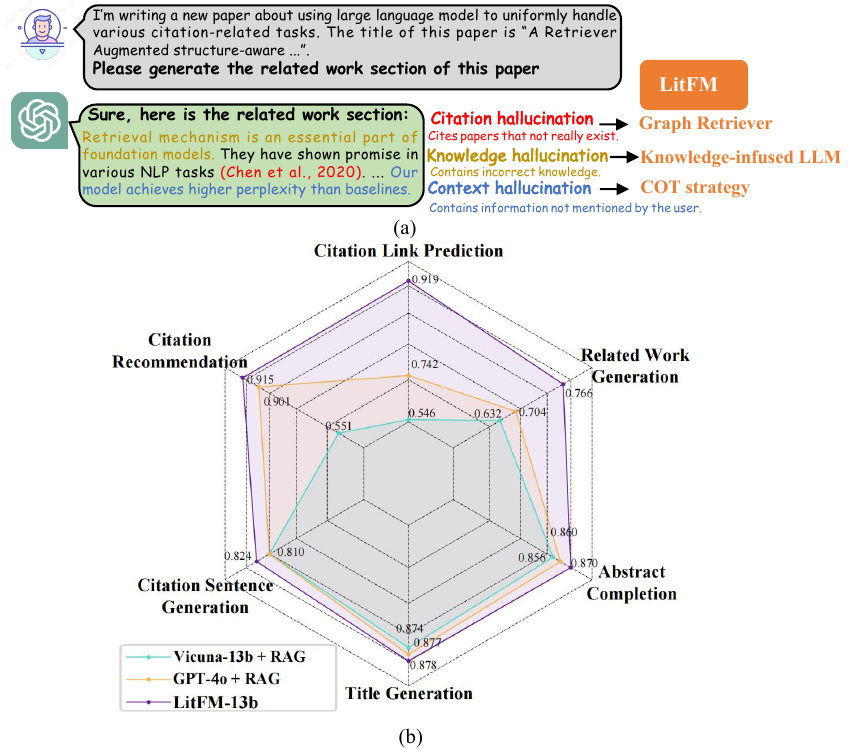}
    \vspace{-10pt}
    \caption{(a) Hallucinations faced by existing approaches. (b) Performance of \method and existing approaches on six benchmark tasks, where \method shows consistent superiority. 
}\vspace{-0.6cm}
    \label{fig:polar}
\end{figure}

Effectively navigating scientific literature to accomplish literature tasks such as citation prediction and related work generation has historically been a labor-intensive process, often requiring manual identification of relevant works through rule-based heuristics \cite{liu2013full}. However, the emergence of large language models (LLMs) has transformed this landscape, enabling these tasks to be tackled in zero-shot settings \cite{zhang2023pre, shen2024citekit, baek2024researchagent}. Despite these advancements, these models tend to generate fabricated content, leading to hallucinations that are particularly problematic in literature tasks. As illustrated in Figure~\ref{fig:polar}(a), this results in responses that exhibit hallucinations in citation, knowledge, and context for a given input~\cite{kunnath2023prompting, martin2024shallow}.

To mitigate knowledge hallucinations, numerous efforts have focused on extending LLMs’ domain-specific knowledge by fine-tuning LLMs on individual literature-related tasks \citep{jung2022intent, martin2024shallow, li2024related}. While these efforts have improved task-specific performance, their narrow task-specific focus overlooks LLMs' ability to transfer knowledge across different tasks \cite{luo2023citationsum, yang2023revisiting}, a limitation that becomes evident in complex applications such as related work generation \cite{li2024related}. Concurrently, recent research has leveraged retrieval-augmented generation (RAG) to ground LLM responses in predefined literature knowledge \citep{gao2023retrieval, zhao2024retrieval}, solving knowledge and citation hallucinations. As literature knowledge is best represented as graphs, these studies have focused on graph-based retrieval over citation graphs \citep{kang2023knowledge, wang2023knowledgpt, shen2023retrieval, jiang2024hykge, he2024g}. Despite providing additional benefits in reducing hallucinations, current state-of-the-art (SOTA) graph-based retrievals face two key limitations: 1) Their performance highly depends on the precision and informativeness of given queries. However, in real-world scenarios, the user-provided queries could be incomplete or ambiguous. As a result, these methods struggle to retrieve the most relevant papers and lead to suboptimal performance. 2) They rely solely on semantic similarity for retrieving relevant papers, which exacerbates the Matthew effect \cite{WANG2014329}—a prevalent bias in which highly cited papers are disproportionately recommended by the retriever, leading to reduced diversity and overlooking potentially relevant literature. Consequently, there remains a critical need for a unified model capable of effectively performing diverse literature tasks—moving beyond a narrow focus on singular tasks—while mitigating hallucination issues, delivering high performance, and addressing the Matthew effect.



To bridge this gap, we propose the first \textbf{Lit}erature \textbf{F}oundation \textbf{M}odel (\method) designed to handle a wide range of literature-related tasks while rivaling the performance of large-scale models. By integrating a state-of-the-art graph retriever specifically tailored for literature graphs and a unified multi-task fine-tuning approach that internalizes citation graph information and its structure, LitFM serves as a literature reasoning agent capable of assisting researchers in real-world applications, such as related work generation. 

Our proposed graph retrieval leverages self-supervised contrastive learning to enhance the model's reliability, particularly in challenging scenarios such as incomplete or ambiguous input queries. Specifically, given an incomplete or ambiguous query, the retriever is trained to reconstruct the golden query in the embedding space (called pseudo-query embeddings), and then use the similarity between the embeddings of the user-provided query and reconstructed pseudo query to enhance the semantic matching. Furthermore, unlike existing graph retrieval methods \cite{he2024g} that rely solely on semantic similarity, our approach incorporates topic guidance to enhance retrieval diversity. By summarizing the user's query into key topics, our graph retriever recommends a diverse set of papers tailored to the target task. Then, to refine the retrieval process, a novel diversity metric is used to re-rank the retrieved papers, effectively addressing the Matthew effect in the process.

To overcome the limitations of single-task approaches and mitigate knowledge hallucinations, we introduce a multi-task instruction-tuning paradigm for citation graph understanding. This framework enables LLMs to internalize both domain-specific knowledge and citation graph structure by unifying diverse literature-related tasks, including both node-level and edge-level tasks, using domain-specific citation subgraphs. Our methodology involves curating citation graphs by processing open-source repositories such as arXiv and PubMed, addressing the lack of citation context in existing datasets \cite{yan2023comprehensive, an2021enhancing}. Then, these curated graphs serve as the foundation for developing comprehensive multifaceted literature tasks datasets. By fine-tuning on these datasets, our model learns to extract intrinsic knowledge, identify key arguments, and capture inter-paper relationships within the target domain. This approach enhances the model's cross-task knowledge transfer capabilities, ultimately creating what we refer to as, a knowledge-infused LLM.

Our evaluation of LitFM demonstrates SOTA performance in its retrieval component, surpassing existing methods on diversity and accuracy. When combined with knowledge-infused LLM, LitFM outperforms leading models (\eg GPT-4o) across literature-related tasks, such as citation link prediction and paper recommendation. To assess its capabilities in complex tasks (\eg related work generation), we design a chain-of-thought (CoT) strategy, characterized by its flexible controllability and ability to mitigate context hallucinations. Results demonstrate that LitFM effectively adapts to domain-specific conventions by aligning its generated related works with the stylistic norms of the target domain. These results are validated through both statistical metrics and comprehensive human evaluation, demonstrating LitFM’s superiority.

\xhdr{Contributions} This paper's contributions are threefold. (1) We propose a novel graph retriever specifically tailored for citation graphs, which captures both semantic and structural relationships and retrieves a more diverse and balanced set for comprehensive literature coverage. (2) LitFM presents a unified multi-task fine-tuning approach that facilitates knowledge transfer across diverse tasks. Experimental results demonstrate that LitFM outperforms state-of-the-art models, including GPT-4, Deepseek-R1, and other existing retrieval-augmented models, across a broad spectrum of literature-related tasks. (3) We open-source LitFM\footnote{\url{https://github.com/zjs123/LitFM}}, along with a graphical user interface, facilitating seamless model usage and enabling broader adoption in research and academic applications.


%% file: chapter/02_relatedwork.tex
\section{Related Work} 

\xhdr{Deep Learning on Citation Graphs}
Citation graphs are extensively utilized in numerous research studies~\cite{zhang2023large}. Early works have applied graph neural networks model~\citep{hamilton2017inductive} for citation graph analysis and recommendation~\citep{iqbal2021decade, ali2020deep}. However, they are often designed for a specific task, lacking a unified model that performs well across different tasks. Efforts such as SciBERT \citep{beltagy2019scibert}, BioBERT \citep{lee2020biobert}, PubMedBERT \citep{gu2021domain}, SciMult \citep{zhang2023pre} aim for general understanding of literature but are only applicable for discriminative tasks due to their emphasis on representation learning. Recent works \citep{jung2022intent, martin2024shallow, li2024related} leverage LLMs to enable generation tasks but are typically designed for specific tasks. Further, they often separate training from real citation graphs, limiting adaptability to network dynamics.

\xhdr{Large Language Models for Graphs}
Existing strategies for leveraging LLMs in graph learning~\citep{minaee2024large, ren2024survey, jin2024large, chen2024exploring} include using GNNs as encoders to convert graph entities into structure-aware tokens for LLM inference~\citep{tang2024graphgpt, chai2023graphllm}, or using LLMs as text encoders for node embeddings to train GNNs~\citep{huang2023prompt, xia2024opengraph, chen2023label} or fused GNN-LLM models~\citep{li2023grenade, huang2024can}. Another branch involves retrieving and translating subgraphs into text for inference~\citep{he2024g, hu2024grag}, with fine-tuning to improve LLM understanding of subgraph context~\citep{zhao2023graphtext, fatemi2023talk, chen2024llaga, luo2024graphinstruct}.


\xhdr{Retrieval Augmented Generation} Vanilla LLMs commonly face challenges like hallicinations~\citep{zhang2023siren}, outdated knowledge~\citep{zhang2023howlarge}, and uninterpretable reasoning process~\citep{jie2024interpretable, tang2024higpt} in domain-specific or knowledge-intensive tasks. RAG addresses these issues by incorporating external knowledge as additional context to improve LLM performance~\citep{gao2023retrieval}. Originating from language applications, most of the current work on RAG for LLMs is on textual data gathered from one or multiple corpuses~\cite {gao2023retrieval, zhao2024retrieval}. Several later attempts adapt these approaches for (semi-) structural data such as tabular data~\citep{luo2023augmented, zha2023tablegpt}, knowledge graphs~\citep{kang2023knowledge, wang2023knowledgpt, shen2023retrieval, edge2024local, jiang2024hykge, luo2025gfm, he2024g, peng2024graph, mavromatis2024gnn}. \citet{he2024g} and \citet{hu2024grag} are the closest work to ours that enhances graph question-answering capabilities by extracting subgraphs based on the retrieved nodes. However, they ignore the semantic-structural dependency during retrieval.

%% file: chapter/03_method.tex
\vspace{-0.3cm}
\section{Methodology}
\xhdr{Problem Formulation} Our overarching goal is to develop a literature foundational model that can effectively address various literature-related tasks commonly encountered in real-world scientific research. These tasks include, for example, generating a title from a paper's abstract, completing an abstract, and creating a related works section specific to domains such as medicine and physics. To achieve this, we define $\boldsymbol{x}$ as the user query related to these tasks. A citation graph $\mathcal{G}=(V,E)$ specific to the domain of interest is available, where each node $i\in V$ represents a paper, and edges $E$ denote citation relationships between papers. \method produces an output $\boldsymbol{y}$ to address the query $\boldsymbol{x}$ by 1) producing distinct query topics according to the user's personalized purposes. 2) navigating $\mathcal{G}$ to extract reference papers, with the pseudo-query embedding addressing cases where $\boldsymbol{x}$ has incomplete information. 3) generating $\boldsymbol{y}$ with a knowledge-infused LLM which aligns domain-specific knowledge through a unified instruction paradigm.

\begin{figure*}
    \centering
    \includegraphics[width=\linewidth]{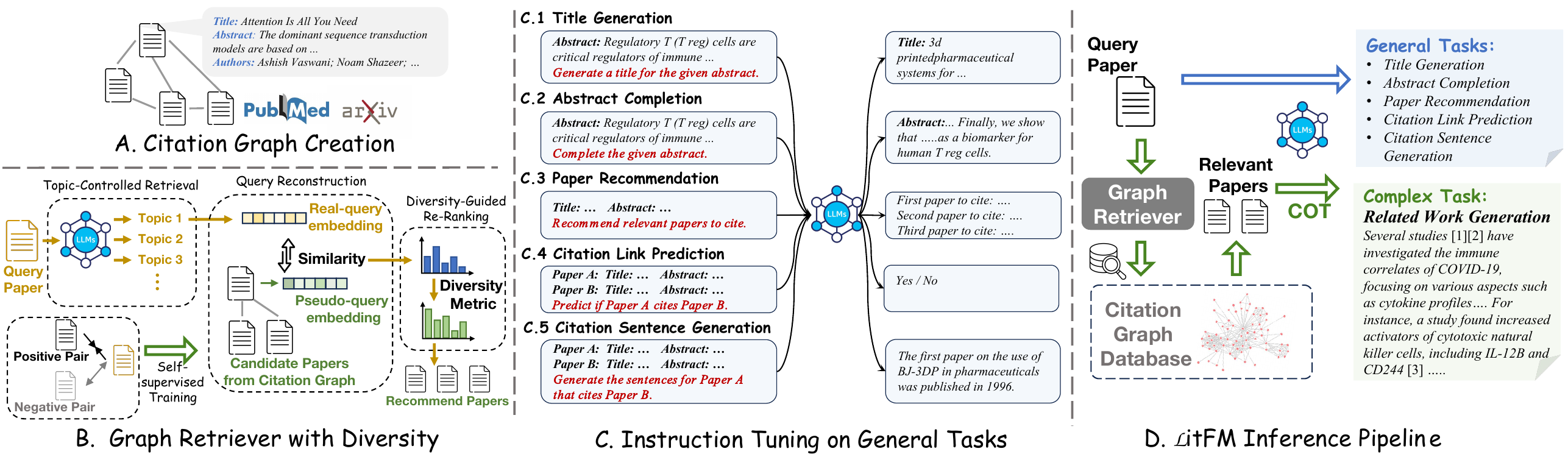}
    \vspace{-0.7cm}
    \caption{Main components of \method. (A) We curate citation graph benchmark datasets with enriched citation context based on which the domain-specific sets for citation instruction tuning are constructed. (B) Graph retriever with query reconstruction. Self-supervised pre-training is employed to adapt to domain properties. The topic-control strategy and the re-ranking are proposed to enhance the retrieval diversity. (C) Instruction paradigm for citation graph understanding, which infuses domain-specific knowledge into LLMs. (D) The retriever-augmented pipeline uniformly manages various literature-related tasks.
    }
    \label{fig:pipeline}\vspace{-0.5cm}
\end{figure*}

\xhdr{Method Overview} Figure \ref{fig:pipeline} illustrates \method's pipeline, which includes the creation of citation graphs for training and evaluation, a graph retriever that balances the retrieval accuracy and diversity, an instruction paradigm to integrate literature knowledge and unify various tasks, and an inference process employing a chain-of-thought strategy to address complex tasks, \eg related work generation. We detail them in the following.

\vspace{-0.2cm}
\subsection{Graph Retriever}
Citation graphs from different domains often exhibit various structural and semantic properties (\eg citation preferences and terminology). Modeling these properties is essential for accurate retrieval. However, existing RAG methods \cite{he2024g} generally do not address these domain-specific aspects and lack diversity in their recommendations, making them unsuitable for personalized retrieval and resulting in lower retrieval quality. 
To address this, our graph retriever is incorporated with controllable retrieval topics and the re-ranking strategy to enhance its recommendation diversity. We propose the novel notion of pseudo-query embedding to generalize the model to incomplete queries. With graph retriever, \method can scale to citation contexts that exceed the input window limitation and extract accurate literature references to reduce hallucinations. It comprises three main steps: indexing, self-supervised training, and retrieval.


\xhdr{Indexing}
We initiate the retriever by generating embeddings for the literature. For a paper $i\in\mathcal{G}$, let $T_i$ and $A_i$ denote its title and abstract. The embedding for node $i$ is defined as $\textbf{z}_i = LM(T_i) + LM(A_i) \in \mathbb{R}^d$, where $d$ denotes the embedding dimension and $LM$ refers to the pre-trained language model, such as BERT~\cite{devlin2019bert}. 

\xhdr{Self-supervised Training}{
Connections in citation graphs indicate the semantic and structural relevance among literature, which is essential for understanding the domain properties. Therefore, we propose a self-supervised training process requiring the model to reconstruct connections between literature. 
Specifically, we first randomly select paper $i$ from 
$\mathcal{G}$ as the query paper and generate the query embedding as $\mathbf{q}_i = \mathbf{W}^q \mathbf{z}_i + \mathbf{b}^q \in \mathbb{R}^{d_1}$, where $\mathbf{W}^q\in\mathbb{R}^{d_1\times d}$ and $\mathbf{b}^q\in\mathbb{R}^{d_1}$ are trainable parameters, and $d_1$ indicates the hidden dimension. Then, the papers with connections to $i$ are regarded as positive candidates and the other ones as negative. The embedding of a candidate paper $j$ is generated by aggregating embeddings of neighborhood papers with one message passing layer~\cite{gilmer2017neural}
\vspace{-0.2cm}
\begin{equation}
    \mathbf{c}_j = \mathbf{W}^{c1} \mathbf{z}_j + \frac{1}{|\mathcal{N}(j)|} \sum_{k \in \mathcal{N}(j)} \mathbf{W}^{c2} \mathbf{z}_k + \mathbf{b}^c \in \mathbb{R}^{d_1},
    \vspace{-0.2cm}
\end{equation}
where $\mathcal{N}(j)$ is the neighborhood papers of $j$. $\mathbf{W}^{(\cdot)}\in\mathbb{R}^{d_1\times d}$ and $\mathbf{b}^c\in\mathbb{R}^{d_1}$ are learnable during training.
Integrating neighborhood information helps the retriever consider the broader context of papers, thereby enhancing the quality and diversity of the retrieval.

In practical usage, user-provided queries may be incomplete (\eg partial titles or fragmented sentences), leading to less-informative queries. We found that for each candidate it is possible to generate a golden query that best matches this candidate. For example, the candidate text ``self-attention mechanism" best matches the query ``the backbone of Transformer architecture". This inspires us to learn to approximate the embedding of the golden query for each candidate. We call the generated one pseudo-query embedding and use the similarity between the embeddings of the user-provided query and pseudo query as an auxiliary metric during retrieval. Since our goal is to predict the potential connections between papers, each pair of connected papers (\ie $(i, j)\in E$) in the citation graph can be regarded as a ground-truth query-candidate match, where the text of the source node $i$ serves as the golden query for the text of the target node $j$. Therefore, we first use a multilayer perceptron (MLP) to generate the pseudo-query embedding from candidate paper $j$ as $\mathbf{p}_j = \text{MLP}(\mathbf{c}_j) \in \mathbb{R}^{d_1}$, and then optimize $\mathbf{p}_j$ to closely resembles the embedding of golden query $i$. Here, we use the $\ell_1$-norm regularization for optimization since it achieves the best performance in our experimental studies. The loss function is
\vspace{-0.2cm}
\begin{equation}\label{eq:regular}
    \mathcal{L}_{\textnormal{re}} = \sum_{(i,j) \in \mathcal{G}} ||\mathbf{z}_i - \mathbf{p}_j||_1.
\vspace{-0.1cm}
\end{equation}

Given the candidate embedding $\mathbf{c}_j$ and the pseudo-query embedding $\mathbf{p}_j$, we calculate the cosine similarity between the query and the candidate with the enhancement of pseudo-query as $\textnormal{sim}(i, j) = \cos(\mathbf{q}_i, \mathbf{c}_j + \mathbf{p}_j)$. We then apply the InfoNCE loss \cite{oord2018representation} to ensure positive candidates are closer to the query embeddings than negative candidates in the semantic space:
\vspace{-0.2cm}
\begin{equation}
    \mathcal{L}_{\textnormal{nce}} = -\frac{1}{|\mathcal{N}(i)|} \sum_{j \in \mathcal{N}(i)} \textnormal{log} \frac{\textnormal{exp}(\textnormal{sim}(i, j))}{\sum_{j' \notin \mathcal{N}(i)} \textnormal{exp}(\textnormal{sim}(i, j'))}, 
\vspace{-0.1cm}
\end{equation}
where the negative candidates $j'$ are randomly sampled from $V\setminus\mathcal{N}(i)$. We then optimize the retriever with the InfoNCE loss $\mathcal{L}_{\textnormal{nce}}$ and regularization term $\mathcal{L}_{\textnormal{re}}$ jointly.

\xhdr{Retrieval}
For each query text $\boldsymbol{x}$ provided by users or generated by LLMs, we create the corresponding query embedding as $\mathbf{q}_x = \mathbf{W}^q LM(\boldsymbol{x}) + \mathbf{b}^q$, where $LM$ indicates a language model as the encoder, \eg BERT~\cite{devlin2019bert}. Then, we use the k-nearest searching approach to retrieve a set of related papers based on the similarity between the query embedding and the candidate embedding, which is accompanied by the corresponding pseudo-query embedding. The retrieval for query text $\boldsymbol{x}$ is defined as:
\vspace{-0.2cm}
\begin{equation}
    V_k(\boldsymbol{x}) = \argtopk_{n \in V} \textnormal{cos}(\mathbf{q}_x, \mathbf{c}_n + \mathbf{p}_n),
\vspace{-0.2cm}
\label{eq:ranking}
\end{equation}
where the $\argtopk$ operation retrieves the top-$k$ elements based on the similarity. $k$ is a hyper-parameter, indicating the number of retrieved papers. Our graph retriever, with its neighbor-aware candidate embedding and structure-aware training, effectively captures structural dependencies that previous methods typically overlook. Additionally, by utilizing pseudo-query embedding, our approach can handle sparse contexts and less informative queries, addressing challenges that previous methods struggle with.


\vspace{-0.2cm}
\subsection{Balancing Retrieval Quality-Diversity} 
\label{sec:Solving Matthew Effects}
While our graph-based retriever effectively predicts likely references, it also presents challenges that could limit a researcher’s exposure to diverse works and inadvertently reinforce the ``Matthew effect'' — where well-established, highly cited papers continue to gain disproportionate attention, making it harder for emerging research to be discovered and recognized. This is because the retriever is trained on real-world citation graphs containing citations across diverse topics. To generalize across topics, it primarily captures coarse-grained similarities between papers rather than fine-grained, subfield-specific relationships. As a result, it tends to recommend broadly related papers rather than precisely tailored recommendations based on a paper’s specific subfield, leading to highly homogeneous recommendations for papers within the same domain. Additionally, due to degree bias \cite{xu2023grace}, where the model inherently favors high-degree nodes, the message-passing mechanism prioritizes highly cited papers, making them more likely to be recommended while limiting the visibility of other relevant but less-cited papers. This amplifies the citation-driven popularity while neglecting others in the research community.

\xhdr{Topic-Controlled Retrieval} To solve these challenges, we propose topic-controlled retrieval to enable diverse and tailored recommendations for different papers. Specifically, rather than directly using the title and abstract as the query text during retrieval, we propose to first use LLM to generate $M$ distinct research topics from the target paper according to user interests. Subsequently, we combine the embeddings of the title, abstract, and text description as the query embedding, which can be defined as 
\vspace{-0.2cm}
\begin{equation}
    \mathbf{z}^m_i = LM(T_i) + LM(A_i) + LM(D^m_i),
\vspace{-0.2cm}
\end{equation}
where $D^m_i$ is the text description of the $m$-th generated topic. $\mathbf{z}^m_i$ is the query embedding for the $m$-th topic of paper $i$. We provide a detailed illustration of the topic generation prompt in Figure \ref{fig:COT}. The integration of topic-controlled retrieval offers three key benefits: 1) It distills the paper’s main arguments from multiple perspectives, ensuring concise retrieval targets and reducing noise from lengthy abstracts, enhancing model performance in general tasks. 2) It enables direct control over the diversity of related work generation. A larger hyper-parameter $M$ allows the model to expand the discussion to more relevant topics, incorporating citations from various aspects. 3) By refining the topic generation prompt with specific instructions, such as focusing on a subfield, our graph retriever adapts to human-driven research needs, enabling tasks like predicting whether one paper will cite another based on their research relationships in the deep learning domain specifically.

\xhdr{Diversity-Guided Re-Ranking} We further propose a diversity-guided re-ranking strategy to mitigate the effects of degree biases. Instead of ranking candidate papers solely by their semantic similarity to query text (as in Eq. \ref{eq:ranking}), we incorporate a diversity measure. Specifically, for candidate papers generated for a topic, we first calculate the central semantic embedding of these papers as
\vspace{-0.25cm}
\begin{equation}
    \mathbf{C}_m = \sum_{n \in \mathcal{T}(m)} LM(T_n) + LM(A_n),
    \vspace{-0.25cm}
\end{equation}

where $\mathcal{T}(m)$ is the candidate papers for topic $m$. Then we get the diversity ranking of papers based on their semantic similarity with the central semantic embedding (i.e., $-\textnormal{cos}(\mathbf{C}_m, LM(T_n) + LM(A_n))$). The higher-ranked paper has lower similarity, making it relatively diverse to the central topic. Each candidate paper now has two rankings: similarity and diversity. We sum them and recommend the top-$k$ papers with the lowest total rank, ensuring a balance between retrieval accuracy and diversity.

\vspace{-0.3cm}
\subsection{Multi-Task Objectives}
\label{instruction_understanding}
We leverage an LLM as the foundation to unify multiple literature-related tasks. To enhance its capabilities on citation graphs and mitigating knowledge hallucinations, we conduct instruction fine-tuning with LoRA~\cite{hu2021lora}, using a curated collection of general tasks on citation graphs. This multi-task instruction tuning approach enables the model to learn task-specific behaviors while preserving general domain-specific knowledge.

As illustrated in Figure \ref{fig:pipeline} (C), we design node-level tasks such as \textit{Title Generation} and \textit{Abstract Completion} based on given paper abstracts. These tasks aim to infuse domain-specific context into the LLM and enhance its scientific writing capabilities. By learning to generate titles and complete abstracts, the model develops a deeper understanding of scientific literature and content representation.

We also incorporate edge-level tasks \textit{Paper Recommendation}, \textit{Citation Link Prediction}, and \textit{Citation Sentence Generation}. They improve the model's comprehension of both scientific context and local structures of citation relations. Specifically, \textit{Paper Recommendation} and \textit{Citation Link Prediction} enable the model to understand the relevance and connections between papers, while \textit{Citation Sentence Generation} helps it learn how citations are integrated into scientific writing, leading to emergent capabilities in handling more complex literature-related queries, such as related work generation. 

By combining node-level and edge-level tasks, our approach ensures a robust understanding of both the content of individual papers and the relations between them. These tasks collectively form a necessary and sufficient set for the model to develop a comprehensive and foundational understanding of citation relations. Detailed instruction formats are provided in Appendix A.

\vspace{-0.4cm}
\subsection{Inference Pipeline}
As shown in Figure~\ref{fig:pipeline} (D), \method can uniformly handle various literature-related tasks. The process begins by retrieving relevant papers from the citation graph, which are then used as references. The LLM then extracts the main arguments and relationships from these references for task-specific generation. 
For example, when determining if paper A would cite paper B, \method first retrieves potential citations using its graph retriever component. The information from two papers and the retrieved references are fed into the knowledge-infused LLM with task-specific instructions. If the arguments of A and B are relevant or they share similar relations with the references, the response will be ``YES". 
While citation graph references may not be needed for tasks like title generation, results in Figure \ref{fig:ablation_analysis} show that our retrieval-augmented pipeline improves performance in tasks like citation recommendation without compromising effectiveness elsewhere. We then extend our inference pipeline to a complex task, related work generation, using a chain-of-thought strategy, offering a more comprehensive evaluation of \method's zero-shot generalization ability.



\xhdr{Chain-of-Thought Strategy}  Generating related work requires more sophisticated synthesis and contextualization of research, going beyond simple retrieval or citation matching. Existing approaches rely on provided references (\eg ground-truth relationships between papers) and lack controllability (\eg cannot personalize the discussion topics). To address this, we introduce a chain-of-thought (COT) strategy, which achieves end-to-end generation with controllability, while mitigating context hallucinations. We refer to Appendix A for more details.
\begin{itemize}[noitemsep,topsep=0pt,parsep=0pt,partopsep=0pt,leftmargin=*]
    \item \textbf{Summarize the query text.} The knowledge-infused LLM first summarizes the main research topics of the paper based on the user-provided title and abstract. The prompt is designed to generate $K$ most relevant topics of this paper that should be discussed in the related work section, where $K$ is a hyper-parameter to control the diversity of the related work. 
    
    \item \textbf{Retrieve related papers.} The generated topics are fed into the graph retriever to retrieve potentially cited papers.

    \item \textbf{Recommend cited papers.} To refine the retrieval, we ask the knowledge-infused LLM to recommend the top-k papers most likely to be cited from the retrieved list.

    \item \textbf{Generate citation sentences.} We ask the LLM to generate the citation sentences that describe the relationship between the target paper and each cited paper,  which are essential for a comprehensive literature review.

    \item \textbf{Group citation sentences as paragraphs.} Given the cited papers and citation sentences for each topic, we ask the LLM to group citation sentences for each topic as a paragraph, allowing similar papers to be discussed together coherently.

    \item \textbf{Combine as a related work section.} Finally, we ask the LLM to combine these paragraphs as a related work section, where the citations should be preserved.
\end{itemize}

\xhdr{Controllability of \method}
Users may have specific preferences for generated related work, such as the number of citations or key discussion topics, depending on their needs. This calls for controllability in related work generation to tailor outputs to individual requirements. However, existing methods either generate related work in a single step \cite{shi2023towards} or rely on retrieved papers without personalized refinement \cite{chen2022target}, offering limited user control.

In contrast, \method offers a more flexible way to improve controllability in related work generation. It assigns specific discussion topics before retrieval, allowing the topic-generation prompt to be customized with precise instructions. For example, for a paper on an LLM-based biology analysis approach, the instruction "Give me $M$ most relevant topics in biology analysis" ensures that retrieved citations focus on biology rather than other LLM applications, effectively steering the discussion in the related work section.

Second, the hyperparameter $M$ effectively controls both the number of paragraphs and the diversity of the related work section. Since the proposed COT strategy organizes papers by topic, a larger $M$ results in more paragraphs, with each addressing a distinct topic, leading to a broader and more diverse discussion. Finally, the number of cited papers is determined by the `Recommend cited papers' step, which selects the top $L$ most relevant papers from retrieved candidate papers. Therefore, the hyperparameter $L$ controls how many papers should be included, which not only refines the cited papers but also provides controllability.

\xhdr{Complexity Analysis}
Compared with vanilla LLMs, the additional time complexity of \method stems from the graph retriever. First, to generate the neighbor-aware candidate embeddings, we perform message passing, which has a complexity of $O(|\mathcal{N}(i)|d)$, where $|\mathcal{N}(i)|$ indicates the number of neighbor nodes for any node $i\in\mathcal{G}$. Then, during inference, the graph retriever encodes the prompt $\boldsymbol{x}$ using a BERT model, which has a time complexity of $O(|x|^2d)$, where $d$ is the embedding dimension and $|\cdot|$ indicates the token length. Finally, the similarity calculation has a time complexity of $O(d_1)$, where $d_1$ represents the hidden dimension.


%% file: chapter/04_experiment.tex
\vspace{-0.5cm}
\section{Experiments}
\subsection{Dataset} 
To establish our framework, we curated three new citation graph benchmark datasets from three distinct domains: medicine, computer science, and physics. Compared with existing citation graph datasets such as ogbn-arxiv-TA \cite{yan2023comprehensive} and  Semantic Scholar Network dataset \cite{an2021enhancing}, our datasets offer several advantages:
\begin{itemize}[noitemsep,topsep=0pt,parsep=0pt,partopsep=0pt,leftmargin=*]
    \item Besides the title and abstract, node attributes in our graph include the paper's related work section (when available).
\item Each edge is annotated with the citing sentence extracted from the citing paper, along with its local context (\ie the preceding and following sentences), thus capturing how the papers discuss one another. Additionally, each edge contains an indicator specifying whether the cited paper is mentioned in the related work section of the citing paper, thus distinguishing between literature review citations and in-paper technical citations.
\end{itemize}
Our dataset not only facilitates the construction of the dataset for instruction tuning but also allows for the evaluation of downstream tasks detailed in the previous sections and serves as a test bed for assessing the performance of the related work generation task. The medicine portion is constructed based on the PubMed Central repository. Meanwhile, the computer science and physics categories are derived from the LaTeX sources from the arXiv repository. For dataset statistics, examples, and more detailed information about the construction process, please refer to Appendix B.

\vspace{-0.4cm}
\subsection{Evaluation on Literature-related Tasks}

\xhdr{Baselines} For generation tasks, we use Mistral 7B \cite{jiang2023mistral}, Vicuna 7B/13B \cite{vicuna2023}, Llama-3 8B\cite{llama}, GPT-3.5-turbo, and GPT-4o as baselines. We also employ a typical GNN-based model, GAT \cite{velivckovic2018graph}, as the baseline for citation link prediction and recommendation tasks. To integrate text information, we initialize the node embedding with the BERT-encoded embeddings of the paper title and abstract. The edge embeddings are initialized by the text of citation sentences. We also include recent models for citation graph understanding, including SciLitLLM \cite{li2024scilitllm}, SciRIFF \cite{wadden2024sciriff}, and OAG-BERT \cite{liu2022oag}.

\xhdr{Experimental Settings} For each citation graph, we randomly sample a connected dense sub-graph with 2,000 nodes as the test set. The remaining nodes and edges, excluding those in the test set, constitute the training data used for training the graph retriever and fine-tuning the large language models via the LoRA technique \cite{hu2021lora}. We use Vicuna model as the LLM backbone of \method. We use the BERT-base-uncased model to yield text embeddings for GNN and graph retriever. For the citation link prediction task, the ratio of positive to negative samples is set to $1:1$. For the citation recommendation task, the size of the candidate set is fixed at $10$ for each query with both negative samples and candidate samples randomly sampled. For the abstract completion task, the first 10\% of each abstract is provided as input to the LLM. The number of retrieved papers that are used for augmentations is set as $k \in \{1, 3, 5, 10\}$. We run all the models five times with different random seeds and report the average performance to eliminate deviations. All the experiments are conducted on NVIDIA A40 with 48 GB memory. More details can be found in Appendix C.


\xhdr{Evaluation Metrics} We use Accuracy for the citation link prediction task and Hits@$k$ for the citation recommendation task. For generation tasks, we follow the previous setting~\cite{zhang2019bertscore} and use the BERTScore to evaluate the semantic similarity of the generated text and the ground truth one. We also measure the relevance of the generated related work using ROUGE-L score~\cite{lin2004rouge}. Moreover, we report the statistics, such as text length and number of citations, in the ground-truth related work and the generated text.


\begin{table}[t]
\centering
\caption{Performance on the citation link prediction task and the citation recommendation task. We report accuracy for the link prediction and the Hits@1 metric for the recommendation task. The best and the second-best results are shown in \textbf{bold} and \underline{underlined} respectively.
}
\vspace{-0.3cm}
\label{tab:text_generation}
\resizebox{0.9\linewidth}{!}{
\begin{NiceTabular}{l|ccc|ccc}\toprule
 & \multicolumn{3}{c|}{\textbf{Citation Link Prediction}} & \multicolumn{3}{c}{\textbf{Citation Recommendation}} \\
 \midrule
 Datasets & \multicolumn{1}{c}{CS} & \multicolumn{1}{c}{Physics} & \multicolumn{1}{c|}{Medicine} &\multicolumn{1}{c}{CS} & \multicolumn{1}{c}{Physics} & \multicolumn{1}{c}{Medicine} \\ \midrule
GAT + BERT & 0.865 & 0.871 & 0.882 & 0.837 & 0.866 & 0.857 \\
\midrule
Vicuna-7b & 0.522 & 0.503 & 0.519 & 0.595 & 0.614 & 0.451 \\
Vicuna-13b & 0.551 & 0.505 & 0.557 & 0.475 & 0.567 & 0.504 \\
Mistral-7b & 0.578 & 0.517 & 0.561 & 0.625 & 0.638 & 0.552  \\
Llama3-8b & 0.615 & 0.522 & 0.559 & 0.683 & 0.682 & 0.617 \\
GPT 3.5 turbo & 0.689 & 0.617 & 0.641 & 0.793 & 0.845 & 0.736 \\
GPT 4o & 0.728 & 0.711 & 0.803 & \underline{0.883} & 0.897 & 0.862 \\
\midrule
SciLitLLM & 0.847 & 0.805 & 0.775 & 0.632 & 0.728 & 0.748 \\
SciRIFF & 0.517 & 0.585 & 0.793 & 0.592 & 0.601 & 0.776  \\
OAG-BERT & 0.790 & 0.679 & 0.721 & 0.713 & 0.735 & 0.743  \\
\midrule
\method-7b & \underline{0.889} & \underline{0.895} & \underline{0.896} & 0.872 & \underline{0.901} & \underline{0.885} \\
\method-13b & \textbf{0.903} & \textbf{0.919} & \textbf{0.953} & \textbf{0.906} & \textbf{0.915} & \textbf{0.908} \\
\midrule
Improve & 4.3\% & 5.5\% & 8.0\% & 8.2\% & 5.6\%  & 5.9\%  \\
\bottomrule
\end{NiceTabular}}
\vspace{-0.5cm}
\label{tab:LP_tasks}
\end{table}

\begin{table}[]
\caption{Performance of existing LLMs and \method on generation tasks with Precision and F-score of the BERT score.}
\vspace{-0.3cm}
\centering
\resizebox{0.9\linewidth}{!}{
\begin{NiceTabular}{c|c|cc|cc|cc} 
\toprule 
\textbf{Models}
&\textbf{Datasets}
&\multicolumn{2}{c}{\textbf{Title Generation}}
&\multicolumn{2}{c}{\textbf{Abstract Completion}}
&\multicolumn{2}{c}{\textbf{Citation Sentence Generation}}\\
\midrule 
&\textbf{Metric}&{Precision}&{F-score}&{Precision}&{F-score}&{Precision}&{F-score}\\
\midrule 
\multirow{3}{*}{Vicuna-7b}&CS &0.875 &0.887  &0.873 &0.852 &0.782  &0.796 \\
 &Physics  &0.866 &0.867  &\underline{0.883} &0.854 &0.829  &0.805 \\
 &Medicine &0.870 &0.875  &0.877 &0.848 &0.845  &0.832 \\

\midrule

 &CS &0.875 &0.888  &0.882 &0.854 &0.792  &0.809 \\
 Vicuna-13b &Physics  &\underline{0.866} &0.868  &0.882 &0.853 &0.829  &0.801 \\
 &Medicine  &0.872 &0.877  &0.880 &0.854 &0.834  &0.825 \\

\midrule

 \multirow{3}{*}{ Mistral-7b}&CS &0.869 &0.872  &0.872 &0.850 &0.825  &0.809 \\
 &Physics  &0.827 &0.854  &0.875 &\underline{0.861} &0.819  &0.798 \\
 &Medicine &0.859 &0.855  &0.872 &0.840 &0.857  &0.833 \\

\midrule

 \multirow{3}{*}{Llama3-8b} &CS &0.841 &0.866  &0.867 &0.838 &0.821  &0.831\\
  &Physics  &0.832 &0.855  &0.846 &0.851 &0.820  &0.799 \\
 &Medicine &0.868 &0.879  &0.879 &0.849 &0.833  &0.827 \\

 \midrule

  \multirow{3}{*}{GPT-3.5 turbo}&CS &0.869 &0.885  &\textbf{0.877} &0.854 &0.829  &0.816 \\
  &Physics &0.862 &0.873  &0.878 &0.855 &0.826  &0.801 \\
 &Medicine &0.862 &0.873  &0.877 &0.849 &0.823  &0.825 \\

 \midrule

  \multirow{3}{*}{GPT-4o}&CS &\underline{0.875} &\textbf{0.889}  &0.857 &0.851 &0.832  &0.824 \\
 &Physics &0.864 &\underline{0.875}  &0.864 &0.853 &\underline{0.834}  &0.808 \\
 &Medicine &\textbf{0.889} &\underline{0.884}  &0.879 &0.854 &\underline{0.860}  &0.831 \\

 \midrule

  \multirow{3}{*}{SciLitLLM}&CS &0.819 &0.827  &0.858 &0.849 &0.815  &0.812 \\
 &Physics &0.851 &0.856  &0.841 &0.835 &0.791  &0.785 \\
 &Medicine &0.863 &0.877  &0.871 &0.865 &0.842  &0.829 \\

 \midrule

  \multirow{3}{*}{SciRIFF}&CS &0.868 &0.871  &0.869 &0.857 &0.817  &0.823 \\
 &Physics &0.857 &0.860  &0.848 &0.853 &0.801  &0.797 \\
 &Medicine &0.851 &0.855  &0.865 &0.858 &0.831  &0.836 \\

 \midrule
 \midrule

\multirow{3}{*}{\method-7b } &CS &0.871 &0.880  &0.864 &\underline{0.860} &\underline{0.836}  &\underline{0.838} \\
 &Physics &0.863 &0.871  &0.875 &0.855 &0.829  &\underline{0.818} \\
 &Medicine &0.877 &0.881  &0.887 &0.851 &0.849  &\underline{0.836} \\

  \midrule

 \multirow{3}{*}{\method-13b}&CS &\textbf{0.880} &\textbf{0.889}  &\underline{0.873} &\textbf{0.863} &\textbf{0.849}  &\textbf{0.841} \\
  &Physics  &\textbf{0.872} &\textbf{0.878}  &\textbf{0.891} &\textbf{0.870} &\textbf{0.837}  &\textbf{0.824} \\
 &Medicine &\underline{0.885} &\textbf{0.890}  &\textbf{0.892} &\textbf{0.859} &\textbf{0.863}  &\textbf{0.849} \\
\bottomrule
\end{NiceTabular}}
\label{tab:generation_tasks}\vspace{-0.5cm}
\end{table}

\xhdr{Performance Comparison} As shown in Table~\ref{tab:LP_tasks} and~\ref{tab:generation_tasks}, \method consistently outperforms other baselines on five evaluation tasks across the three datasets, especially for tasks requiring graph comprehension. On the medicine dataset, \method achieves 8.0\% improvement in accuracy for citation link prediction and 5.9\% in Hits@1 for citation recommendation task, highlighting the effectiveness of our proposed model in reducing hallucinations and filtering out irrelevant information. Additionally, \method achieves the best performance on the citation sentence generation task across all three datasets, showing its superiority in reasoning relationships between literature. Notably, \method-7B version outperforms LLMs with larger parameter sizes, such as Llama3-8b and Vicuna-13b. Interestingly, \method shows a more significant improvement on the medicine dataset, which can be attributed to 1) the corpus in this dataset containing many medical terms, thus requiring more domain-specific knowledge, and 2) the sparsity of this dataset, which benefits more from the neighbor-aware modeling and pseudo-query embedding. 

\xhdr{Ablation and Analysis}
\begin{figure}[t]
\vspace{-0.2cm}
\centering
\subfigure[]{\includegraphics[width=0.45\linewidth]{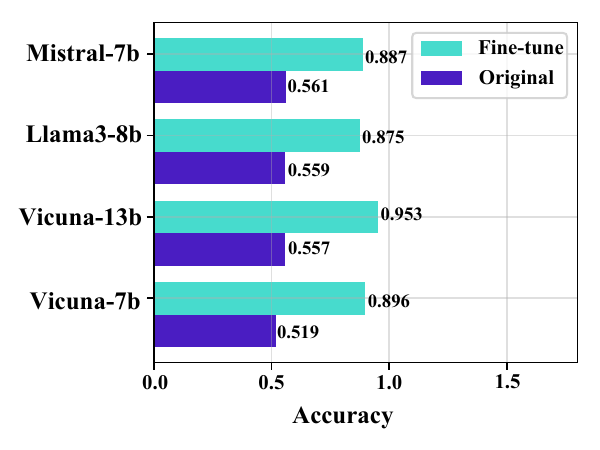}}
\subfigure[]{\includegraphics[width=0.45\linewidth]{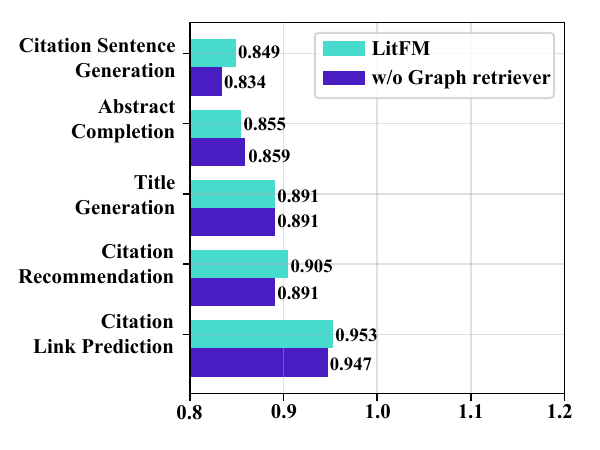}}\vspace{-0.3cm}
\caption{(a) Link prediction performance of existing LLMs with and without knowledge-infused fine-tuning. (b) Performance of \method with and without graph retriever.} 
\label{fig:ablation_analysis}\vspace{-0.3cm}
\end{figure}
As illustrated in Figure~\ref{fig:ablation_analysis} (a), our instruction tuning strategy and the graph-augmented pipeline can be seamlessly adapted to other LLMs to enhance their performance on citation graph tasks, which shows the generalization ability of our proposed strategies. In Figure~\ref{fig:ablation_analysis} (b), we observe that the graph retriever enhances performance across most tasks, especially those requiring graph comprehension. For title generation and abstract completion tasks, which require an understanding of the semantics within the paper, the integration of additional information from other papers does not lead to performance degradation, showing the robustness of \method in capturing relevant information.

\subsection{Graph Retriever Analysis}

\begin{table}[t]
\centering
\caption{Retrieval performance of different approaches. P@5 and P@10 refer to Precision@5 and Precision@10.}
\vspace{-0.3cm}
\label{tab:text_generation}
\resizebox{0.9\linewidth}{!}{
\begin{NiceTabular}{l|cc|cc|cc}
\toprule
 \textbf{Datasets} & \multicolumn{2}{c}{\textbf{CS}} & \multicolumn{2}{c}{\textbf{Physics}} & \multicolumn{2}{c}{\textbf{Medicine}} \\
 \midrule
 \textbf{Models} & \multicolumn{1}{c}{P@5} & \multicolumn{1}{c}{P@10} &\multicolumn{1}{c}{P@5} & \multicolumn{1}{c}{P@10} &\multicolumn{1}{c}{P@5} & \multicolumn{1}{c}{P@10} \\ \midrule
G-Retriever~\cite{he2024g} & 0.286 & 0.301 & 0.202 & 0.305 & 0.389 & 0.433 \\
GRAG~\cite{hu2024grag} & 0.309 & 0.343 & 0.208 & 0.326 & 0.403 & 0.467 \\
\midrule
\midrule
\textbf{Graph retriever} & \textbf{0.623} & \textbf{0.760} & \textbf{0.736} & \textbf{0.832} & \textbf{0.723} & \textbf{0.714} \\
w/o pseudo-query embedding & 0.519 & 0.587 & 0.597 & 0.643 & 0.644 & 0.542 \\
w/o neighbor-aware candidate embedding & 0.331 & 0.357 & 0.220 & 0.141 & 0.351 & 0.379 \\
\bottomrule
\end{NiceTabular}}
\label{tab:retrieval_accuracy}
\end{table}

\begin{figure}[t]
\vspace{-0.5cm}
\centering
\subfigure[]{\includegraphics[width=0.45\linewidth]{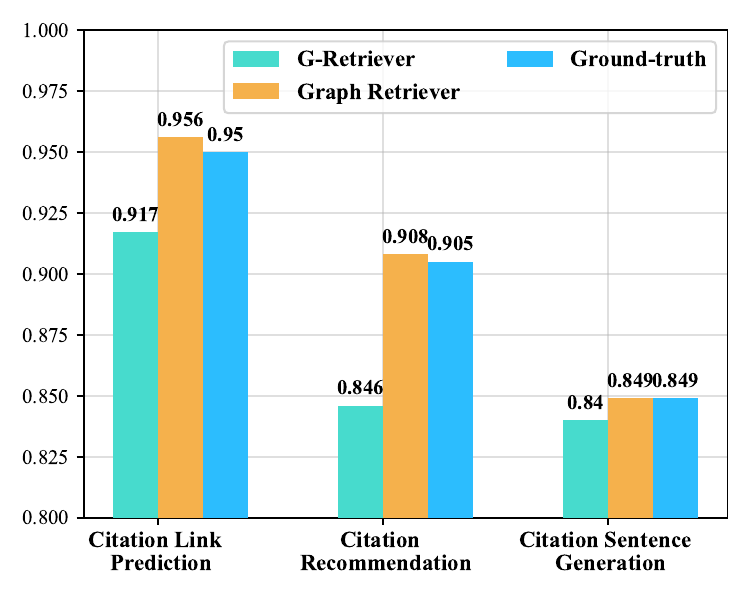}}
\subfigure[]{\includegraphics[width=0.45\linewidth]{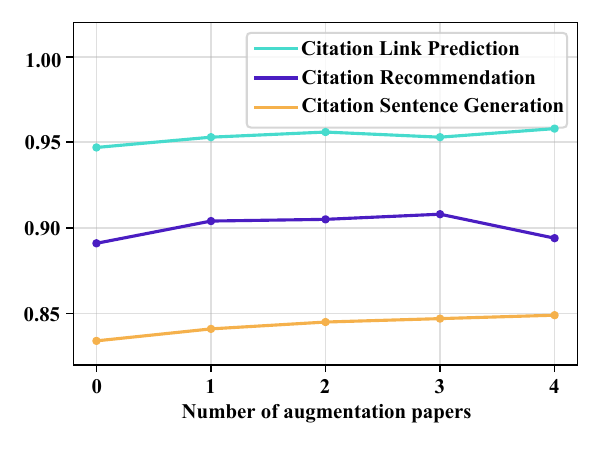}}\vspace{-0.3cm}
\caption{(a) Performance of \method when using different retrieving approaches. (b) The performance of \method with different number of augmentation papers.} 
\label{fig:retriever_analysis}
\end{figure}
\begin{figure}[t]
\centering
\vspace{-0.5cm}
{\includegraphics[width=0.9\linewidth]{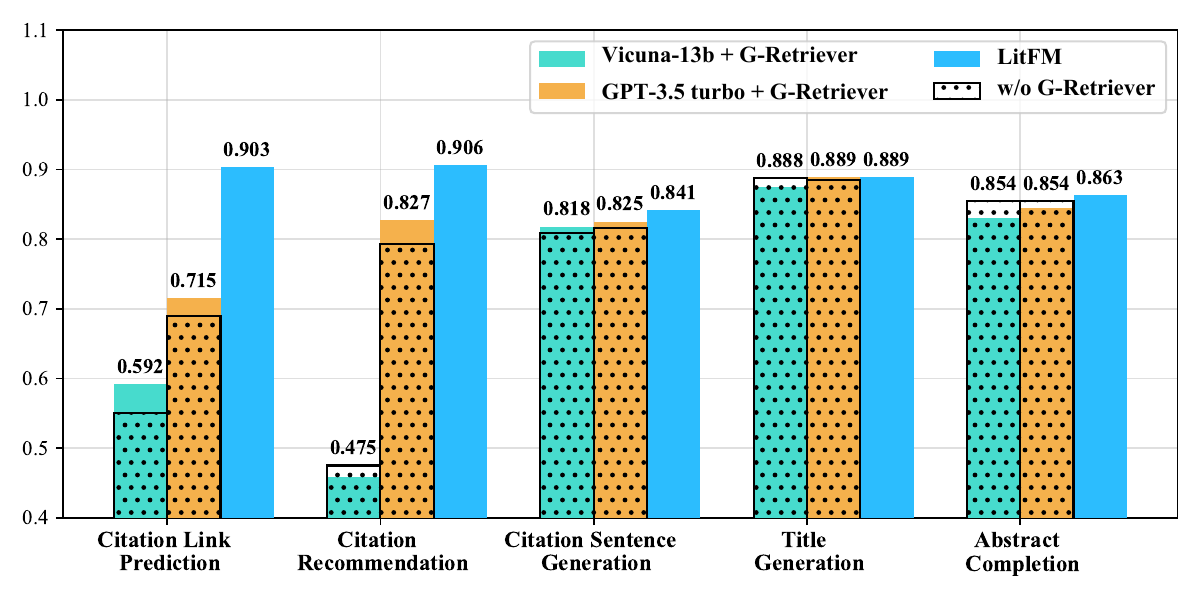}}\vspace{-0.3cm}
\caption{Performance of existing LLMs with G-Retriever on citation graph tasks.} 
\label{fig:vicuna_GPT_G_retriever}\vspace{-0.5cm}
\end{figure}

\xhdr{Retrieval Accuracy} To analyze the effectiveness of our proposed graph retriever, we compare its retrieval precision with typical graph-aware RAG approaches including G-Retriever \cite{he2024g} and GRAG \cite{hu2024grag}, and two variants of our graph retriever: 1) without pseudo-query embedding, and 2) without neighbor-aware candidate embedding. As shown in Table \ref{tab:retrieval_accuracy}, the graph retriever achieves higher retrieval precision compared with existing methods, especially for the sparse physics dataset, showing the effectiveness of the proposed pseudo-query embedding and neighbor-aware candidate embedding in handling sparse context scenarios. Notably, even without pseudo-query embedding, \method still consistently outperforms baselines, highlighting the efficacy of neighbor-aware training in capturing the structural context of papers.

\xhdr{Augmentation Effectiveness} Figure \ref{fig:retriever_analysis} (a) shows the performance of \method with different
retrieving approaches. We observe that \method achieves higher generation quality with our graph retriever, especially for the citation link prediction task where inaccurate retrieval can introduce noise text and thus damage the performance. This shows the effectiveness of the graph retriever in accurately enriching citation context. One interesting observation is that using references retrieved by our graph retriever leads to higher accuracy on link prediction task than using the ground-truth references. This can be attributed to the graph retriever's ability to filter out papers that are not directly related. Figure \ref{fig:retriever_analysis} (b) illustrates the performance trend as the number of augmentation papers increases. While a larger number of augmentation papers enhances the quality of generated citation sentences, it leads to a slight decline in citation recommendation performance due to the noise introduced by irrelevant papers. Figure \ref{fig:vicuna_GPT_G_retriever} demonstrates that \method consistently outperforms the retrieval-augmented Vicuna and GPT-3.5 turbo. Vicuna exhibits performance degradation on recommendation and title generation with G-Retriever. This could be attributed to Vicuna's lack of domain-specific knowledge, which hinders its ability to extract valuable information from the augmentation paper.

\begin{table}[t]
\centering
\caption{Retrieval diversity performance. Distribution means the distribution distance between real-world and recommended citation papers. Similarity measures the semantic similarity of recommended papers. Overlap indicates the proportion of identical recommendations for different papers within the same topic. $^{\ast}$ indicates the proposed graph retriever used in LitFM.}
\vspace{-0.3cm}
\resizebox{0.85\linewidth}{!}{
\begin{NiceTabular}{l|c|c|c}
\toprule
 \textbf{Datasets} & \multicolumn{1}{c}{\textbf{Distribution}} & \multicolumn{1}{c}{\textbf{Similarity}} & \multicolumn{1}{c}{\textbf{Overlap}} \\
 \midrule
bert-base-uncased~\cite{he2024g} & 142.19 & 0.877 & 0.246 \\
BAAI~\cite{he2024g} & 131.76 & 0.875 & 0.133 \\
G-Retriever~\cite{he2024g} & 135.77 & \underline{0.859} & 0.132 \\
\midrule
\midrule
Graph retriever$^{\ast}$ & \underline{127.46} & 0.868 & \underline{0.117}\\
Graph retriever-D$^{\ast}$ & \textbf{123.51} & \textbf{0.837} & \textbf{0.093}\\
\bottomrule
\end{NiceTabular}}
\label{tab:diversity}
\end{table}

\xhdr{Matthew Effect} In this part, we take the CS graph as an example to show our graph retriever can mitigate the Matthew effect and enhance the recommendation diversity. 

First, the retriever should mitigate degree bias and align the recommendation with real-world citation distributions. As shown in Table \ref{tab:diversity}, we measure the distribution distance between real-world citations and recommendations from different retrieval models. Graph retriever-D refers to graph retriever enhanced by diversity ranking discussed in Section \ref{sec:Solving Matthew Effects}. The results show that our graph retriever better aligns with real-world distributions, while topic-controlled retrieval and re-ranking further improve this alignment.

Second, ideally, the retrieved papers for each topic should align with the broad topic while presenting diverse key arguments and contributions. This ensures a more comprehensive discussion in the related work section. As shown in Table \ref{tab:diversity}, we measure the semantic similarity of recommended papers for each retrieval model. Our graph retriever achieves better semantic diversity with the enhancement of the re-ranking strategy.

Finally, previous retrieval models relied on coarse-grained semantic matching, often recommending the same papers for papers within the same broad topic. Ideally, recommendations should be diverse, reflecting each paper’s key arguments and contributions. As shown in Table \ref{tab:diversity}, we measure the proportion of identical recommendations for different papers within the same topic. Our graph-based retriever, leveraging topic-controlled retrieval, better captures each paper’s specific arguments, reducing recommendation overlap.

\subsection{Related Work Generation Performance}
\begin{table}[t]
\caption{Performance of existing LLMs and \method on related work generation tasks.}
\vspace{-0.3cm}
\centering
\resizebox{0.9\linewidth}{!}{
\begin{NiceTabular}{c|cc|cc} 
\toprule 
\textbf{Datasets}
&\multicolumn{2}{c|}{\textbf{CS}}
&\multicolumn{2}{c}{\textbf{Physics}}\\
\midrule 
\textbf{Models}
&\multicolumn{1}{c}{BERT Score}
&\multicolumn{1}{c|}{ROUGE}
&\multicolumn{1}{c}{BERT Score}
&\multicolumn{1}{c}{ROUGE}\\
\midrule 
Llama3-8b  &0.765 &0.158 &0.671 &0.168 \\

Vicuna-13b  &0.735 &0.191 &0.632 &0.162 \\

GPT-3.5 turbo  &0.801 &0.190 &0.684 &0.170 \\

GPT-4o  &\underline{0.806} &\underline{0.221} &\underline{0.697} &\textbf{0.217} \\

 \midrule
 \midrule
 
\method-13b &\textbf{0.819}$^{(\uparrow1.6\%)}$&\textbf{0.224}$^{(\uparrow1.3\%)}$ &\textbf{0.766}$^{(\uparrow9.8\%)}$ &\underline{0.177} \\
\bottomrule
\end{NiceTabular}}
\label{tab:related_work_performance}\vspace{-0.3cm}
\end{table}

\xhdr{Performance Comparison} As shown in Table \ref{tab:related_work_performance}, \method outperforms existing powerful LLMs on related work generation on both BERT and ROUGE score. This demonstrates that the related work section produced by \method is more semantically similar to the ground truth and covers more information from the cited papers. Table \ref{tab:related_work_performance_statistic} shows that the statistical distributions of \method's generated content are more aligned with the ground truth, especially for the sparse physics dataset, highlighting the superior generalization ability of \method. Although GPT-4o generates longer related work sections, it often fails to segment paragraphs effectively. In contrast, \method, benefiting from its knowledge-infused LLM, can better understand literature semantics and achieve more rational segmentation.  Moreover, although the related work generated by existing LLMs includes some citations, these cited papers may not correspond to real papers. Instead, \method can refer to actual citation graphs, thus reducing the citation hallucinations.

\begin{table}[t]
\caption{Statistics Comparison between the generated related work and the ground truth one on the average length (L), average number of paragraphs (NP), average number of citations (NC), and average ratio of paragraphs with citations (RPC). The statistics that are closest to the ground truth are shown in \textbf{bold} and the secondary are \underline{underlined}. }
\vspace{-0.3cm}
\centering
\resizebox{\linewidth}{!}{
\begin{NiceTabular}{c|cccc|cccc} 
\toprule 
\textbf{Datasets}
&\multicolumn{4}{c|}{\textbf{CS}}
&\multicolumn{4}{c}{\textbf{Physics}}\\
\midrule 
\textbf{Models}
&\multicolumn{1}{c}{L}
&\multicolumn{1}{c}{NP}
&\multicolumn{1}{c}{NC}
&\multicolumn{1}{c|}{RPC}
&\multicolumn{1}{c}{L}
&\multicolumn{1}{c}{NP}
&\multicolumn{1}{c}{NC}
&\multicolumn{1}{c}{RPC}\\
\midrule 
Ground Truth &1014.08 &7.36 &21.44 &0.859 &701.43 &6.44 &16.84 &0.689 \\
\midrule 
Llama3-8b  &\underline{378.01} &12.15 &\underline{9.53} &\underline{0.915} &452.90 &16.02 &8.87 &0.874 \\
Vicuna-13b  &214.84 &3.69 &3.77 &0.918 &249.64 &\underline{3.67} &6.53 &0.951 \\
GPT-3.5 turbo  &307.05 &\textbf{6.65} &4.73 &0.736 &202.38 &2.63 &5.36 &0.822 \\
GPT-4o  &\textbf{701.11} &13.98 &9.20 &0.781 &\underline{503.97} &11.59 &\underline{10.04} &\textbf{0.706} \\
 \midrule
 \midrule
\method-13b &338.90 &\underline{5.95} &\textbf{9.57} &\textbf{0.802} &\textbf{504.81} &\textbf{5.81} &\textbf{10.53} &\underline{0.821} \\
\bottomrule
\end{NiceTabular}}
\label{tab:related_work_performance_statistic}\vspace{-0.5cm}
\end{table}

\xhdr{Ablation Study} Table \ref{tab:related_work_ablation} examines the effectiveness of each component of \method in related work generation. The results show COT improves generation quality by breaking down the complex related work generation task into several simple tasks, such as citation recommendation and citation sentence generation. This step-by-step strategy helps \method to accurately extract useful information from the citation graphs. The table also reveals that performance degrades significantly without the graph retriever, especially on the ROUGE score. This decline occurs because, without the augmentation of retrieved papers, the generation often contains many fake citations due to LLM hallucinations. G-Retriever, which fails to consider structural relevance among papers, tends to introduce noise information. Instead, our graph retriever enhances generation quality by utilizing the structural context of papers.

\begin{table}[t]
\caption{Related work generation performance of variants. 
}
\vspace{-0.3cm}
\centering
\resizebox{0.9\linewidth}{!}{
\begin{NiceTabular}{c|cc|cc} 
\toprule 
\textbf{Datasets}
&\multicolumn{2}{c|}{\textbf{CS}}
&\multicolumn{2}{c}{\textbf{Physics}}\\
\midrule 
\textbf{Models}
&\multicolumn{1}{c}{BERT Score}
&\multicolumn{1}{c|}{ROUGE}
&\multicolumn{1}{c}{BERT Score}
&\multicolumn{1}{c}{ROUGE}\\
\midrule 
w/o COT  &0.803 &0.204 &0.621 &0.104 \\
w/o Graph retriever  &0.792 &0.127 &0.616 &0.096 \\
with G-Retriever~\cite{he2024g}  &0.798 &0.176 &0.631 &0.164 \\
 \midrule
 \midrule
 \method-13b &\textbf{0.819 }&\textbf{0.224} &\textbf{0.766} &\textbf{0.177}\\
\bottomrule
\end{NiceTabular}}
\label{tab:related_work_ablation}\vspace{-0.3cm}
\end{table}

\begin{figure*}[t]
    \centering
    \includegraphics[width=0.78\linewidth]{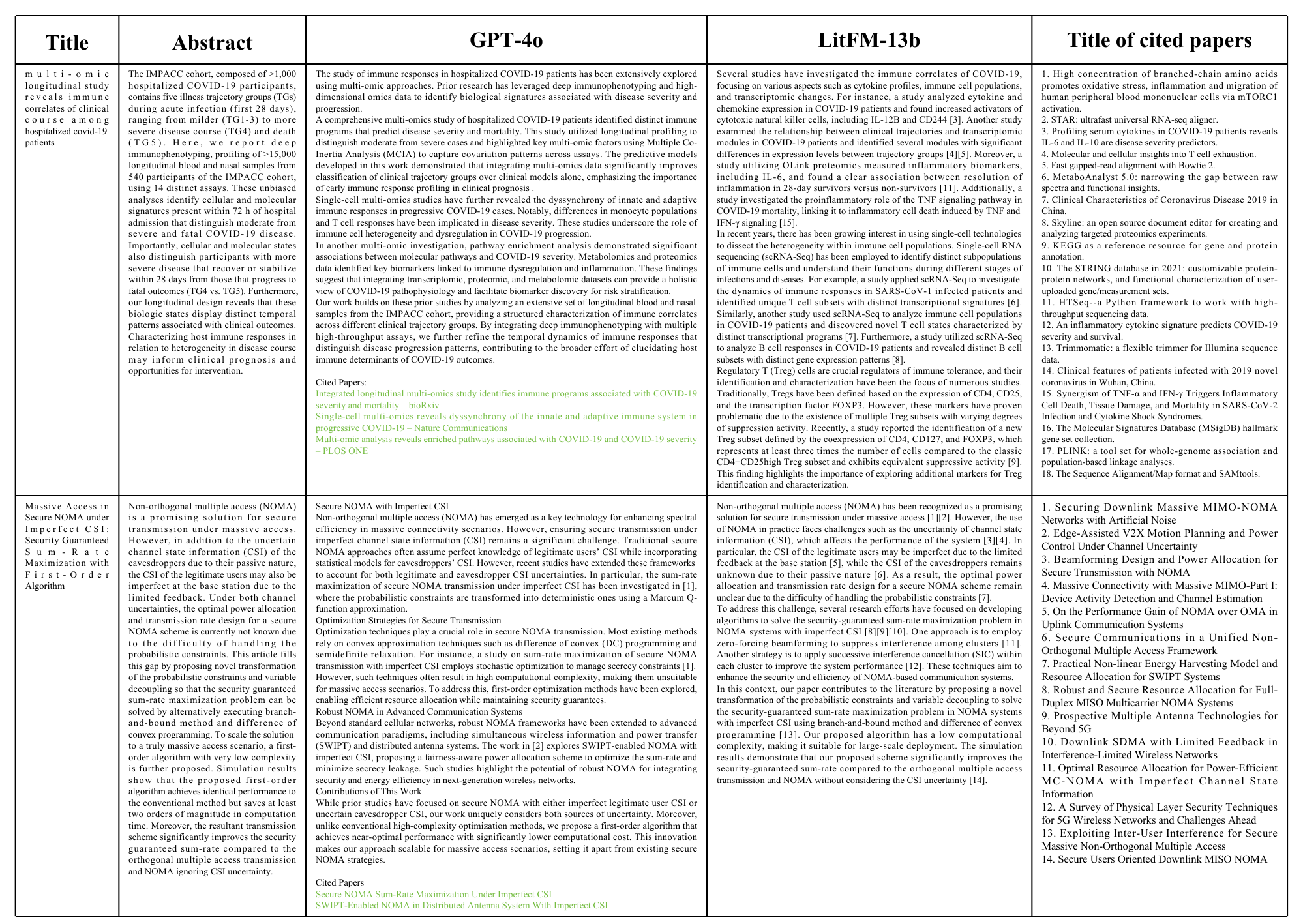}
    \vspace{-3mm}
    \caption{Generation examples on the related work generation task. The green-colored text indicates fake citations.}
    \label{fig:case_related_work}
    \vspace{-3.5mm}
\end{figure*}

\xhdr{Human Evaluation:}
We conducted a human evaluation study using Qualtrics involving several active researchers in the AI domain. Each researcher received excerpts from related work sections generated by LitFM, GPT-4o, and DeepSeek-R1 for 10 different papers. The researchers were instructed to score each model on a scale of 1 to 5, based on several criteria, reported below: \textbf{Q1:} The relevance of the related work to the title and abstract. \textbf{Q2:} The diversity of the cited material. \textbf{Q3:} The clarity of the related work section in terms of structure (division into paragraphs, etc.) and readability. \textbf{Q4:} The level of detail of the related work section. \textbf{Q5:} Rate how appropriate the number of citations is. \textbf{Q6:} Rate how similar the related work section is to others found in this academic field.

\begin{table}[t]
\caption{Human evaluation on related work generation task.}
\vspace{-3mm}
\centering
\resizebox{0.6\linewidth}{!}{
\begin{NiceTabular}{c|cccccc} 
\toprule 

\textbf{Models}
&\multicolumn{1}{c}{Q1}
&\multicolumn{1}{c|}{Q2}
&\multicolumn{1}{c}{Q3}
&\multicolumn{1}{c}{Q4}
&\multicolumn{1}{c}{Q5}
&\multicolumn{1}{c}{Q6}\\
\midrule 
DeepSeek-R1  & \textbf{5} &	4	&4.6 &	3.6&	3.6	&3.2  \\

GPT-4o  &\textbf{5}	&4	&\textbf{5}	&3.6	&3.6	&3.6\\

 \midrule
 \midrule

\method-13b & \textbf{5} &	\textbf{4.8} &	4.2	&\textbf{4.6}&	\textbf{4.8}	&\textbf{4.4} \\
\bottomrule
\end{NiceTabular}}
\label{tab:survey}
\vspace{-7mm}
\end{table}

Researchers were told to give a relative score; a score of 5 means that the model performed best on that particular question. Note that multiple models could receive the same score if their performance was comparable. The results are reported in Table \ref{tab:survey}. As can be seen, the results demonstrate a clear conclusion: LitFM outperforms the other models in terms of topic diversity, suitability to the related work generation style in the target field, and the depth of detail provided alongside citations. LitFM achieved an aggregate score of 4.72, compared to 3.96 for GPT-4o and 3.88 for DeepSeek-R1 on these metrics. However, one area where LitFM lags behind is English structure and coherence, with a score of 4.2, compared to 4.6 for DeepSeek-R1 and 5 for GPT-4o. This is due to the significantly larger number of parameters in GPT-4o and DeepSeek-R1, leading to advanced proficiency in English compared to LitFM.


\xhdr{Case Study} 
As illustrated in Figure \ref{fig:case_related_work}, \method has several advantages in generating literature reviews compared with existing powerful LLM such as GPT-4o. Firstly, It contains more domain-specific words such as ``scRNA-Seq'' and ``zero-forcing beamforming'', showing that \method contains more domain-specific knowledge. Moreover, \method can generate related work sections with explicit citations, and all cited papers are extracted from the citation graph. However, the generation of GPT-4o contains fake citations. Finally, \method shows more reasonable paragraph segmentation compared with GPT-4o. This is because our COT strategy has explicitly asked the \method to group different papers based on their topic and thus achieves a better summary of these cited papers.
\section{Conclusion}
In this paper, we develop the first literature foundation model \method that uniformly handle various citation graph-related tasks. It consists of a graph retriever and a knowledge-infused LLM to avoid hallucinations and generalize to handle new papers outside the citation graph. Experimental results demonstrate both the versatility of \method and its superior performance across six benchmark tasks.

%% file: chapter/05_appendices.tex
\appendix
\section{Instruction Paradigm}
\label{appd:insturction}

\begin{figure*}
    \centering
    \includegraphics[width=0.9\linewidth]{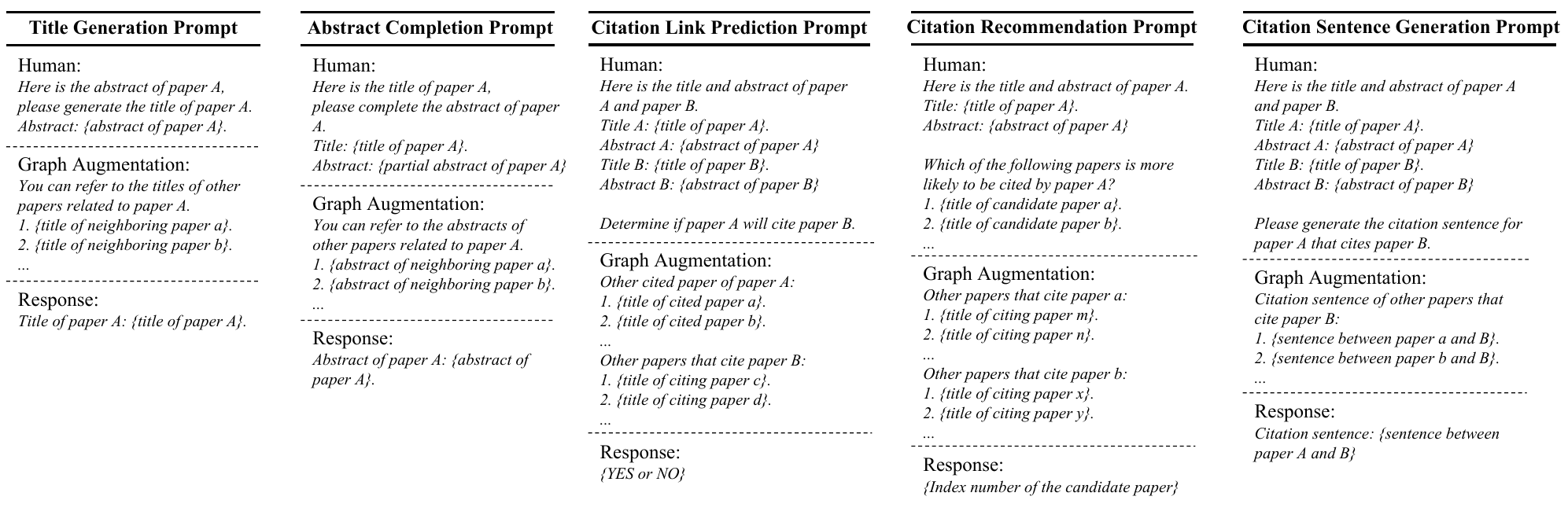}
    \vspace{-0.3cm}
    \caption{Instruction prompt for fine-tuning tasks}
    \label{fig:prompts}
    \vspace{-10pt}
\end{figure*}

\begin{figure*}
    \centering
    \includegraphics[width=0.9\linewidth]{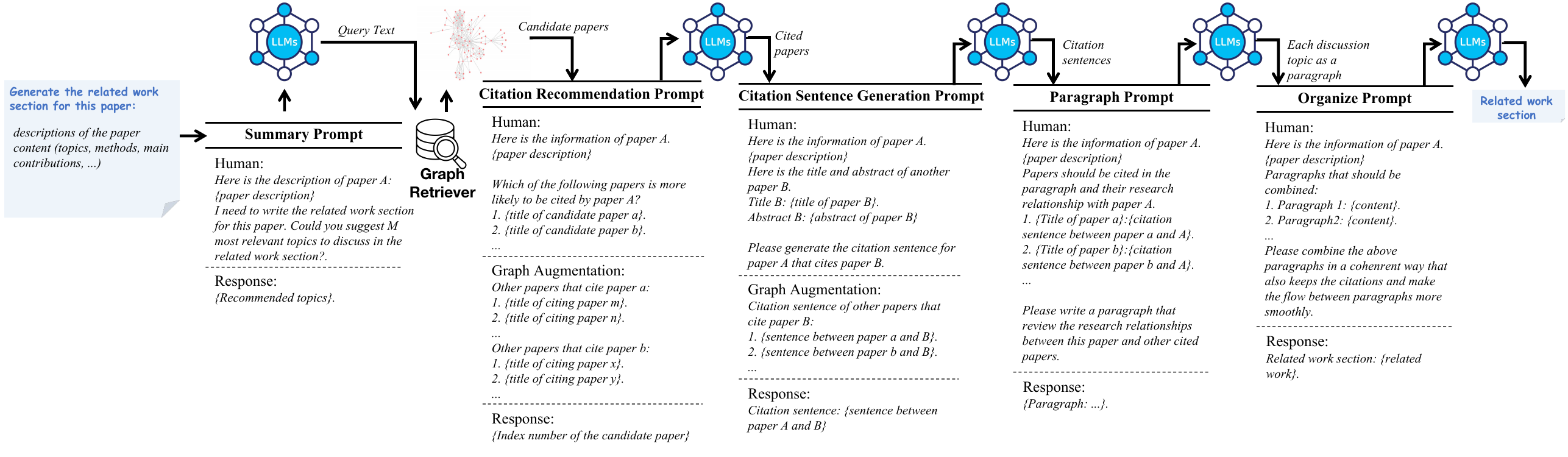}
    \vspace{-0.3cm}
    \caption{Illustration of the chain-of-chain strategy used for relation work generation.}
    \label{fig:COT}
    \vspace{-10pt}
\end{figure*}

As shown in Figure \ref{fig:prompts}, we provide the detailed instruction prompt for each fine-tuning task. The instruction prompt can be seamlessly instantiated through the domain literature and used to inject domain-specific knowledge into the LLM. As illustrated in Figure \ref{fig:COT}, we provide detailed chain-of-thought instructions.

\section{Dataset Details}
\label{sec:appendix_a}
\subsection{Extraction Procedure}
\subsubsection{Arxiv Papers} 
To construct citation graphs for the CS and Physics categories, we first clean the LaTeX source by: (1) removing comments, (2) flattening files, (3) replacing user-defined macros with standard LaTeX commands, (4) removing native LaTeX commands, and (5) standardizing citation syntax. We then parse the bibliography file (.bib) to map citation keys to arXiv IDs. If .bib is unavailable, we fall back to the .bbl file or extract inline citations from the thebibliography environment.

After obtaining the paper content and citation metadata, we locate each citation, extract its citing sentence along with the surrounding context (previous and next sentences), and identify the “Related Work” section. This section is matched using predefined titles (e.g., “Related Work,” “Literature Review,” etc.); if no match is found, we default to the “Introduction.” Each citation is annotated with a flag indicating whether it appears in the related work section.



\subsubsection{PubMed papers} 
For PubMed Central papers, we start by discarding XML files without PMIDs. We then unify citation styles, converting superscripts and in-text names to bracketed references. Citations are extracted via regular expressions and mapped to PMIDs. As with arXiv papers, we extract citation contexts and identify the related work section using the same title-matching strategy, defaulting to the introduction when needed. Each citation is similarly flagged based on its section location.

\begin{table}[t]
\centering
\caption{Statistics of the citations graphs.}
\vspace{-0.3cm}
\resizebox{0.8\linewidth}{!}{
\begin{tabular}[t]{lccc}
\toprule
& \#Nodes & \#Edges & \#Related Works\\
\midrule
Medicine & 2.1M &	7.4M	&1.5M 	 \\
Computer Science & 340k&	3.2M&	188k	 \\
Physics & 59k	& 120k &	19k \\
\bottomrule
\end{tabular}}
\label{table:datastatistics}
\vspace{-10pt}
\end{table}
\subsection{Examples} 
The statistics of our datasets are provided in Table \ref{table:datastatistics}. We further provide below an example of a node and an edge. The [...] symbol is inserted below to reduce the size of the text.

\textbf{Node ID:} 36234872

\textbf{Title:} Influence of the Epoxy/Acid Stoichiometry on the Cure Behavior and Mechanical Properties of Epoxy Vitrimers 

\textbf{Abstract:} Bisphenol A epoxy resin cured with a mixture of dimerized and trimerized fatty acids is the first epoxy vitrimer and has been extensively studied. [...]
properties of epoxy vitrimers can be tuned with the variation in the epoxy/acid stoichiometry. \\
\textbf{Related Work:} Introduction Epoxy resin, one of the most important and popular classes of thermosetting polymers, is firmly rooted in many areas, both in our daily lives and in industry [1,2,3,4,5]. 
[...]
To investigate the effect of recycling on the solvent stability, thermal and mechanical properties of the original epoxy vitrimers, solvent immersion, dynamic mechanical analysis (DMA) and uniaxial tests were performed. \\
\textbf{Edge ID:} (`36234872', `28757995') \\
\textbf{Sentence:} To solve these problems, a thrust to develop a novel class of thermosetting polymers with recyclable, healable and reprocessable features has been carried out during the last two decades. \\
\textbf{Preceding Sentence:} Thus, most of the epoxy wastes are disposed of by landfill or incineration, which has caused not only a waste of resources but also environmental pollution. \\
\textbf{Following Sentence:} In 2011, Leibler et al developed a new class of polymers, called vitrimer, based on associative covalent adaptable networks of epoxy resins cured with a mixture of dimerized and trimerized fatty acids through transesterification reactions [13]. \\
\textbf{In Related Work:} True ~
\textbf{Category:} \color{niceblue} \textbf{Medicine}
\color{black}

\section{Implementation Details}
\label{sec:appendix_c}
\xhdr{Fine-tune LLM}
We construct \method-7b and \method-13b by fine-tuning vicuna-7b and vicuna-13b on domain-specific literature through our proposed institution paradigm. During fine-tuning, the LLMs are loaded in 8-bits and the rank of LoRa is set as 8. Only the $\mathbf{Q}$, $\mathbf{K}$, $\mathbf{V}$, and $\mathbf{O}$ matrices within a large language model will be updated during fine-tuning. We set the batch size to 2 with the maximum number of update steps to 40000, and the gradient accumulation step to 8. The learning rate is set as 0.0002 and we use AdamW \cite{loshchilov2017decoupled} for optimization. The scaling hyperparameter $lora\_alpha$ is set to 32, and the dropout rate during fine-tuning is set to 0.05. To construct the training instruction set for each citation graph dataset, we randomly sample 20,000 nodes from the citation graph and use the title and abstract part of each paper to construct the title generation instructions and the abstract completion instructions. We extract the induced subgraph of these nodes to construct other instructions. Specifically, each edge in the induced subgraph will be used to construct a positive citation link prediction instruction. Then, we replace the target node of the edge with a random node to construct a negative citation link prediction instruction. The ratio of positive to negative samples is set to 1:1. Each citation sentence generation instruction is constructed based on the citation sentence of each edge as well as the titles and abstracts of the connected nodes. For citation recommendation instructions, we randomly sample 10 negative nodes for each edge to construct the candidate set. These instructions will be gathered and shuffled during fine-tuning.

\xhdr{Graph retriever}
We train the graph retriever with Adam \cite{kingma2014adam} optimizer. The number of training epochs is set as 500 and we use the early stopping strategy with a patience of 5.
To integrate the text information, we use the BERT-base-uncased model to get the representations of text attributes, and then these representations are used to initialize the embeddings of nodes and edges. The dimensions of the pre-trained representations are 768 and the maximum text length is set as 512. The BERT model is frozen during training. We set the number of negative samples and the maximum number of neighbor nodes as 10. We use random sampling to generate negative samples. We randomly split the edges in citation graphs into train/validation/test sets by 7:1.5:1.5 and select the best model based on the Precision@5 metric on the validation set.

\xhdr{Baselines}
During inference, the baseline LLMs share the same instruction form as our \method. For all of them, we set the temperature as 0.7 and the nucleus sampling size (i.e., $top\_p$) is set as 0.95. The repetition penalty is set as 1.15 to discourage the repetitive and redundant output. The maximum number of tokens that the LLM generates is set as 1024. We ran all the models five times with different random seeds and reported the average performance.